\documentclass[fleqn,usenatbib]{mnras}

\usepackage{newtxtext,newtxmath}

\usepackage[T1]{fontenc}

\usepackage{graphicx}	
\usepackage{amsmath}	
\usepackage{bm}
\usepackage{comment}
\usepackage{pdflscape}
\usepackage{xspace}
\usepackage[dvipsnames]{xcolor}

\newcommand{\HA}{H$\alpha$\xspace}
\newcommand{\OII}{[\ion{O}{ii}]\xspace}
\newcommand{\Msun}{M_{\sun}}

\newcommand{\hMpc}{h^{-1} \, \mathrm{Mpc}}

\title[Mock ELGs with DM-only Simulations]
{Mock catalogues of emission line galaxies based on the local mass density in dark-matter only simulations}

\author[K. Osato et al.]{
Ken Osato$^{1,2,3}$\thanks{E-mail: ken.osato@yukawa.kyoto-u.ac.jp},
Takahiro Nishimichi$^{1,4}$,
and Masahiro Takada$^{4}$
\\
$^{1}$Center for Gravitational Physics,
Yukawa Institute for Theoretical Physics, Kyoto University,\\
Kitashirakawa Oiwakecho, Sakyo-ku, Kyoto 606-8502, Japan\\
$^{2}$LPENS, D\'epartement de Physique, \'Ecole Normale Sup\'erieure,
Universit\'e PSL, CNRS, Sorbonne Universit\'e, Universit\'e de Paris,\\
24 rue Lhomond, 75005 Paris, France\\
$^{3}$Institut d'Astrophysique de Paris, Sorbonne Universit\'e, CNRS, UMR 7095,
98bis boulevard Arago, 75014 Paris, France\\
$^{4}$Kavli Institute for the Physics and Mathematics of the Universe,
The University of Tokyo Institutes for Advanced Study,\\
5-1-5 Kashiwanoha, Kashiwa-shi, Chiba 277-8583, Japan\\
}

\date{Accepted XXX. Received YYY; in original form ZZZ}

\pubyear{2021}

\begin{document}
\label{firstpage}
\pagerange{\pageref{firstpage}--\pageref{lastpage}}
\maketitle

\begin{abstract}
The high-precision measurement of spatial clustering of emission line galaxies (ELGs)
is a primary objective for upcoming cosmological spectroscopic surveys.
The source of strong emission of ELGs is nebular emission
from surrounding ionized gas irradiated by
massive short-lived stars
in star-forming galaxies.
As a result, ELGs are more likely to reside in newly-formed halos
and this leads to a nonlinear relation between
ELG number density and matter density fields.
In order to estimate the covariance matrix of cosmological observables,
it is essential to produce many independent
realisations to simulate ELG distributions
for large survey volumes.
To this end, we present a novel and fast scheme
to populate ELGs in
dark-matter only $N$-body simulations based on local density field.
This method enables fast production of mock ELG catalogues suitable for
verifying analysis methods and quantifying observational systematics in
upcoming spectroscopic surveys
and can populate ELGs
in moderately high-density regions even though the halo structure cannot be resolved
due to low resolution.
The power spectrum of simulated ELGs is consistent with results of hydrodynamical simulations
up to fairly small scales ($\lesssim 1 h \, \mathrm{Mpc}^{-1}$), and
the simulated ELGs are more likely to be found in filamentary structures,
which is consistent with results of semi-analytic and hydrodynamical simulations.
Furthermore, we address the redshift-space power spectrum of simulated ELGs.
The measured multipole moments of simulated ELGs clearly exhibit a weaker Finger-of-God effect
than those of matter due to infalling motions towards halo centre, rather than random virial motions inside halos.
\end{abstract}

\begin{keywords}
large-scale structure of Universe -- cosmology: theory -- methods: numerical
\end{keywords}



\section{Introduction}
\label{sec:introduction}
The large-scale structure of the Universe is a consequence of the gravitational amplification of
the tiny primordial fluctuations, generated during an inflationary epoch,
in an expanding universe \citep[e.g.][]{2003moco.book.....D}.
In the structure formation dark matter and dark energy play crucial roles;
dark matter causes the gravitational instability and
dark energy sources the cosmic accelerated expansion.
Hence, properties of large-scale structure, measured via the distribution of matter and galaxies,
can be used to probe the cosmic expansion history and constrain cosmological parameters \citep{2013PhR...530...87W}.
Ongoing and upcoming wide-area spectroscopic galaxy surveys  aim at making
a three-dimensional  (more exactly, redshift and two-dimensional angular positions) map of galaxies
over wide solid angle and large redshift leverarm.
Among different tracers, star-forming, emission line galaxies (hereafter ELGs) are
the main targets of upcoming galaxy surveys such as the
Subaru Prime Focus Spectrograph\footnote{\url{https://pfs.ipmu.jp/intro.html}} \citep[PFS;][]{Takada2014},
the Dark Energy Spectrograph Instrument\footnote{\url{https://www.desi.lbl.gov}} \citep[DESI;][]{DESI2016a,DESI2016b},
the ESA \textit{Euclid}\footnote{\url{https://sci.esa.int/web/euclid}} \citep{Amendola2013}
and the NASA Roman Space Telescope\footnote{\url{https://roman.gsfc.nasa.gov}}, because
spectral features (e.g. $4000 \, \text{\AA}$ break) in early-type galaxies become inaccessible
in optical wavelength range for galaxies at $z\gtrsim 1.2\text{--}1.4$ and,
on the other hand, emission lines in star-forming galaxies (e.g. \OII) are relatively easier to identify at such high redshifts.
The baryon acoustic oscillations, measured from the large-scale clustering of galaxy distribution,
are among the most powerful method of cosmological distances that are
least affected by galaxy bias uncertainties \citep{2005ApJ...633..560E,2005MNRAS.362..505C,2021PhRvD.103h3533A}.
In addition, the redshift-space distortion \citep[RSD;][]{Kaiser1987} due to peculiar velocities of galaxies can be
used to test properties of gravity
on cosmological scales \citep{2001Natur.410..169P,2014MNRAS.443.1065B,2016PASJ...68...38O,2021PhRvD.103h3533A}.
Upcoming galaxy surveys promise to make significant improvements in these cosmological constraints.

However, the nature and properties of ELGs involve complicated physics. Most of ELGs are associated with star-formation activities
in the host galaxy \citep{Byler2017},
where massive OB stars ionize the surrounding gas and then the ionized gas emits emission lines such as \HA and \OII lines.
On the other hand, some ELGs can be sourced by active galactic nuclei \citep{Comparat2013}.
Hence it is still difficult and challenging to accurately model the formation and
evolution of ELGs in the current standard structure formation model (the $\Lambda$CDM model).
The most appropriate method is a cosmological hydrodynamical simulation as represented by
the IllustrisTNG simulations \citep{Nelson2019} where the simulation takes into account key physics
such as gas cooling and heating, star formation, supernovae and AGN feedback in subgrid processes \citep[see][for a review]{Somerville2015}.
However, running such simulations is computationally very expensive.
A cosmological analysis quite often requires a large number of mock catalogs each of which is required to cover a sufficiently
large volume, more than $1 \, (h^{-1} \, \mathrm{Gpc})^3$,
to have sufficient statistics of baryonic acoustic oscillation scales \citep[e.g.][]{2020PhRvD.101b3510K}.
Meeting both requirements of resolution or dynamical range
and large cosmological volume for hydrodynamical simulations is still not attainable with current numerical resources.
Hence we need an alternative route to generating desired mocks of ELGs
in a cosmological volume in preparation for upcoming galaxy surveys.
One way is to use a high-resolution, large-volume $N$-body (gravity only) simulation,
e.g. a simulation with trillion particles
in a volume of a few $(h^{-1} \, \mathrm{Gpc})^3$ size \citep{2019ApJS..245...16H,2021ApJS..252...19H,Ishiyama2021},
which can resolve down to dark matter halos with mass of  $\sim 10^{11} \, h^{-1} \, M_\odot$
that likely host ELGs targeted by upcoming surveys \citep{Gonzalez-Perez2018,Gonzalez-Perez2020,2021MNRAS.502.3599H}.
Then the semi-analytical model (SAM) or the halo occupation distribution method can be used to populate ELGs into halos
in the simulation realization. However, having a sufficient number of mock realizations is still difficult
because the original high-resolution $N$-body simulation is already computationally expensive.
Another way is a hybrid method combining low-resolution gravity-only simulation and high-resolution (hydrodynamical) simulation,
assisted with a machine-learning method \citep[e.g.][]{2019arXiv190205965Z,2020arXiv201006608L}.
This method appears to be very promising, but the method requires a careful calibration/training
of the hyper-parameters for each cosmological model,
and thus it is not clear whether one hybrid-method recipe, calibrated for one particular cosmology,
can be used for different cosmological models.

Hence the purpose of this paper is to develop
an even easier method to general a mock catalog of ELGs,
using a low-resolution $N$-body simulation without resolving small halos
($\mathcal{O}(10^3)$ particles with a volume of $\mathrm{Gpc}^3$ scale),
as an alternative route to the aforementioned computational needs for upcoming galaxy surveys.
Our method is designed to identify ``places'' in the cosmic web where ELGs
likely
form,
motivated by the expectation that ELGs tend to form in filaments or sheets of cosmic web,
rather than in nodes or voids \citep{2012MNRAS.423.2617T,Gonzalez-Perez2018,Gonzalez-Perez2020,Orsi2018}.
This is a distinct property of ELGs compared to luminous early-type galaxies
that tend to reside around the center of massive host halos,
where the star formation activity is quenched and the galaxy appears to be red (therefore, no emission line).
In our method, we will propose to use the local mass density at each $N$-body particle position,
estimated by the smoothed hydrodynamical kernel, and then identify such places where ELGs likely reside,
from $N$-body particles satisfying the tuned range of the mass density
so that the distribution of the selected $N$-body particles can fairly well reproduce environment dependence of ELGs
in the IllustrisTNG simulation.
We also study properties of mock ELGs in our method:
radial profiles of ELGs in host halos and real- and redshift-space power spectra of ELGs
compared to those of dark matter or massive subhalos.
We are aware that this method cannot be perfect, but gives an easy method of generating many mocks of ELGs
from low-resolution large-volume $N$-body simulations. The mock ELG catalogues generated in our method can be used to
test/calibrate analysis methods (measurement and parameter inference) and covariance matrix,
including observational effects such as effects of fibre collision and masks/survey geometry \citep{2020JCAP...06..057S}
for a wide range of scales $10^{-3} \lesssim k/[h \, \mathrm{Mpc}^{-1}]\lesssim 1$.

This paper is organised as follows.
In Section~\ref{sec:methods}, we describe the method of generating a mock ELG catalog from a $N$-body simulation.
In Section~\ref{sec:results}, we validate our method by comparing properties of the mock ELGs in our method with those in
the hydrodynamical simulation IllustrisTNG,
and investigate the clustering properties of ELGs.
We conclude in Section~\ref{sec:conclusions}.
Throughout this paper, we assume the flat-geometry $\Lambda$ cold dark matter model
characterized by the following cosmological parameters:
the baryon density parameter $\Omega_\mathrm{b} = 0.0486$, the matter density parameter $\Omega_\mathrm{m} = 0.3089$,
the Hubble constant $H_0 = 100 \, h \, \mathrm{km} \, \mathrm{s}^{-1} \, \mathrm{Mpc}^{-1}$
with $h = 0.6774$, the tilt parameter of primordial curvature power spectrum $n_\mathrm{s} = 0.9674$,
and the present-day root mean square of matter fluctuations at $8 \, h^{-1} \, \mathrm{Mpc}$, $\sigma_8 = 0.8159$.

\section{Methods}
\label{sec:methods}

\subsection{Populating ELGs based on local density}
\label{sec:scheme}
In this section, we describe the method to populate ELGs into
dark-matter only $N$-body simulations.
Our method is designed to identify places in the cosmic web where ELGs likely form or reside,
according to the \textit{local matter density}.
Hence we first compute the local density $\rho_i$
based on the smoothed particle hydrodynamics \citep[SPH;][]{Lucy1977,Gingold1977,Springel2010b} kernel estimation:
\begin{equation}
  \rho_i = \sum_{j} m_\mathrm{p} W_\mathrm{SPH} (r_{ij}/h_i),
\end{equation}
where the subscript $i$ denotes the label of the particle,
$m_\mathrm{p}$ is the particle mass, $r_{ij}$ is the distance between 
particle $i$ in consideration and another particle $j$,
$h_i$ is the smoothing length determined so that the effective number of particles within the smoothing length should be close to $64$,
and the summation runs over particles closer to the particle $i$ than $h_i$.
The kernel function $W_\mathrm{SPH} (x)$ is given as
\begin{equation}
  W_\mathrm{SPH} (x) = \frac{8}{\pi h^3}
  \begin{cases}
    1 - 6 x^2 +6 x^3 , & (0 \leq x < 1/2) \\
    2 (1-x)^3 , & (1/2 \leq x < 1) \\
    0 . & (1 \leq x)
  \end{cases}
\end{equation}
In order to populate ELGs, we select particles which reside in the local density in
the \textit{target} range $\Delta_\mathrm{min} < \rho_i/\bar{\rho}_\mathrm{m} < \Delta_\mathrm{max}$,
where $\bar{\rho}_\mathrm{m}$ is the mean matter density,
and $\Delta_\mathrm{min}$ and $\Delta_\mathrm{max}$ are the minimum and maximum density thresholds that
we can specify to match to the desired clustering properties of mock galaxies.
To be more precise, the maximum density threshold is introduced because ELGs are not likely to reside
in the highest density regions, such as the central region of massive halos where the star formation would be quenched.
On the other hand, the minimum density threshold is introduced to avoid
too low density regions, i.e., voids.
The local density depends on the resolution of the $N$-body simulations
because the same number of neighbours in different mass resolutions
leads to the different effective smoothing scale in density estimation
\citep[for detailed discussions on numerical convergence of SPH, see][]{Zhu2015}.
Hence, the minimum and maximum thresholds need to be chosen in such a way that
the resulting mock catalog reproduces the desired properties of ELGs (clustering strength and environments)
as we will discuss in detail in Section~\ref{sec:thresholds}.

For velocity, we simply assign particle velocity
to bulk velocity of an ELG.
Though the particle velocity may not be appropriate for central ELGs \citep{Guo2015,Yuan2021},
most of mock ELGs are satellites as seen in number density profile (Section~\ref{sec:radial_profile})
and thus this assignment of velocity is justified.

\subsection{Simulations}
\label{sec:simulation}
SAM and/or hydrodynamical simulations are
powerful tools to directly simulate the formation and evolution of ELGs.
However, the computational cost is expensive and these simulations can be run only for a small volume,
compared to the typical volume covered by ongoing and upcoming cosmological surveys.
In practice, the dark-matter only $N$-body simulations,
which are much faster than hydrodynamical simulations, can cover a sufficiently large cosmological volume,
with reasonable computational cost.
Our method presented in Section~\ref{sec:scheme} can be readily applied to
such $N$-body simulations to generate mock catalogues of ELGs that cover a large volume.

To implement our method, we use a large-volume $N$-body simulation with the following specifications.
We employ \textsc{Gadget-4} \citep{Springel2021} code to simulate the gravitational evolution of
the matter density field and compute the SPH local density for each particle.
We run the simulations with a side length of the simulation box $L = 2 \, h^{-1} \, \mathrm{Gpc}$ and
the number of particles $N = 2048^3$, and hereafter, we refer to this simulation as ``Large''.
The particle mass is $m_\mathrm{p} = 7.98 \times 10^{10} \, h^{-1} \, \Msun$.
In this simulation, only the halos with the mass $\gtrsim 10^{13} \, h^{-1} \, \Msun$ can be resolved
with more than 100 $N$-body particles, meaning that small halos hosting ELGs are not fully resolved.
In our method, we identify $N$-body particles that are in the desired range of the local mass density
$\Delta_\mathrm{min} < \rho_i/\bar{\rho}_\mathrm{m} < \Delta_\mathrm{max}$.
Some of the selected particles are in a halo, while other particles are not.
However, we believe that, if we run a higher mass resolution simulation,
the latter group of particles can be found in smaller mass halos that are not resolved by our original simulation.
Thus our method does not entirely rely on halos, so is different from the standard halo occupation distribution method.
Throughout this paper, we apply our method to the snapshot at redshift $z=1$ and
all results are at redshift $z=1$.

For comparison and validation of our mock catalogs, we use the publicly available
IllustrisTNG simulations \citep{Nelson2019,Pillepich2018,Nelson2018,Springel2018,
Naiman2018,Marinacci2018}, which are based on the moving-mesh code \textsc{AREPO} \citep{Springel2010a}.
Among the IllustrisTNG simulation suite, we use the largest-volume dark-matter only
simulation, TNG300-1-Dark run (hereafter, TNG300-D),
which covers the comoving volume of $(205 \, \hMpc)^3$.
The particle mass is $m_\mathrm{p} = 5.9 \times 10^7 \, h^{-1} \, \Msun$
and thus the simulation contains halos with virial masses $> 10^{10} \, \Msun$, where the lower bound corresponds to
a typical smallest halo hosting ELGs.
Furthermore, we utilise the simulated ELG catalogue created with stellar population synthesis code
\textsc{P\'EGASE-3} \citep{Fioc2019} with full physics TNG300-1 run
\citep[also see][for the method of generating mock ELGs from the IllustrisTNG simulation data]{Osato2021},
which includes various galaxy formation physics.
From this catalogue, we construct \HA and \OII ELG samples by selecting ELGs
from the ranked list of line luminosities with the number density cuts
$1.6 \times 10^{-3} \, ( \hMpc )^{-3}$ for \HA ELGs and $2.6 \times 10^{-3} \, ( \hMpc )^{-3}$ for \OII ELGs.
The corresponding luminosity cuts are $1.10 \times 10^{42} \, \mathrm{erg} \, \mathrm{s}^{-1}$ for \HA ELGs and
$5.01 \times 10^{41} \, \mathrm{erg} \, \mathrm{s}^{-1}$ for \OII ELGs.
These selection criteria are determined so that
the large-scale bias is matched to that of mock ELGs
(see the next paragraph).
The dust attenuation effect is taken into account
for the line luminosities, and thus the values above correspond to ``observed'' luminosities.

For comparison, we also construct a mock luminous red galaxy (LRG) sample.
In principle, LRGs are stellar mass limited sample in contrast to ELGs,
which constitute star formation rate limited sample.
We define the mock LRG samples as a sample of massive subhalos in TNG300-D and Large simulations
which are composed of subhalos with the mass larger than $10^{13} \, h^{-1} \, \Msun$.
Note that (sub)halos are identified with \textsc{Subfind} algorithm \citep{Springel2001}.
The number density of the LRG samples is
$n_\mathrm{LRG} = 1.81 \times 10^{-4} \, (\hMpc)^{-3}$ for TNG300-D and
$n_\mathrm{LRG} = 2.58 \times 10^{-4} \, (\hMpc)^{-3}$ for Large,
which are roughly consistent with a target number density of LRGs
at $z\sim 1$ for the DESI survey \citep{DESI2016a}.

In our model, the particles which reside in the local density in the target range are candidates of ELGs.
Although it is possible
to match to the desired galaxy number density $n_\mathrm{ELG}$ by
downsampling from the candidates,
in order to reduce the shot noise, we use all candidate particles but impose a constant weight $w$
to match the number density:
\begin{equation}
  \label{eq:weight}
  \left. w = n_\mathrm{ELG} V \middle/
  \sum^N_{\Delta_\mathrm{min} < \rho_i/\bar{\rho}_\mathrm{m} < \Delta_\mathrm{max}} 1 , \right.
\end{equation}
where $V$ is the simulation box volume and the summation runs over all particles.
Throughout this paper, we adopt $n_\mathrm{ELG} = 10^{-3} \, (\hMpc)^{-3}$,
which roughly corresponds to a typical number density of ELGs
that upcoming Subaru PFS
($n_\mathrm{ELG} \simeq [2.0,8.0] \times 10^{-4} \, (\hMpc)^{-3}$
for $0.8 < z < 2.4$; \citet{Takada2014}) and
DESI ($n_\mathrm{ELG} \simeq [2.5,7.0] \times
10^{-4} \, (\hMpc)^{-3}$ at
$0.6<z<1.4$; \citet{DESI2016a})
are designed to have.
We adopt $\Delta_\mathrm{max} = 500$ and $\Delta_\mathrm{min} = 50$ for TNG300-D,
$\Delta_\mathrm{max} = 500$ and $\Delta_\mathrm{min} = 2$ for Large as our default choices.
Here, $\Delta_\mathrm{min}$ is determined so that the linear bias parameter of
the selected $N$-body particles becomes $b_1 = 1.72$, which is a typical value of
the linear bias parameter of ELGs as found from mock ELGs in the hydrodynamical simulations (see the next section).
On the other hand, $\Delta_\mathrm{max}$ is more
responsible for small scale dynamics.
Lower $\Delta_\mathrm{max}$ removes more particles
near the centre of halos, where the velocity of particles
is dominated by the virial motion,
and as a result of setting $\Delta_\mathrm{max}$, the smearing effect on anisotropic power spectrum
at small scales, i.e., Finger-of-God (FoG) effect, is more suppressed.
By tuning $\Delta_\mathrm{min}$ and $\Delta_\mathrm{max}$,
we can have desired properties of linear bias parameter and environment dependence.
Thus our method is designed to be simple
and easy to implement with $N$-body simulations.
Hence we can generate a sufficiently large number of the mock realisations with our method, which is one of the main purposes.

\begin{figure}
\includegraphics[width=\columnwidth]{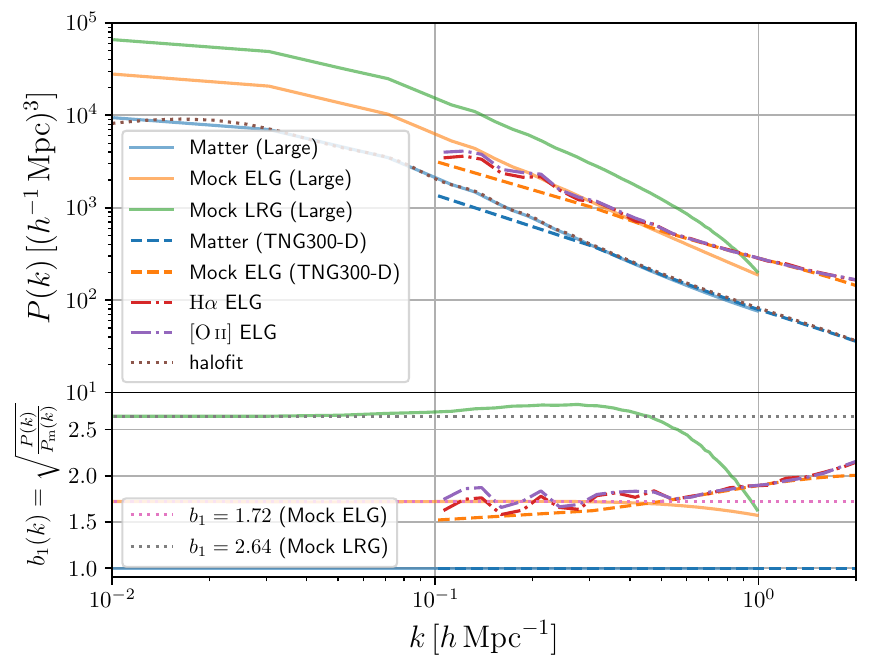}
\caption{Power spectra of mock ELGs, mock LRGs, and matter density distributions at redshift $z = 1$ (upper panel)
and scale-dependent bias of ELG and LRG power spectra (lower panel).
The solid (dashed) line shows the results of the TNG300-D (Large) simulation.
The power spectra of \HA and \OII ELGs generated from IllustrisTNG full physics run are also shown.
The purple dot-dashed line shows the \textit{halofit} result \citep{Takahashi2012} for reference.
In the lower panel, the constant bias ($b_1 = 1.72$ for mock ELGs and $b_1 = 2.64$ for mock LRGs) is also shown as dotted lines.
}
\label{fig:power}
\end{figure}

First, in Figure~\ref{fig:power}, we show
the power spectra of mock ELGs (either \HA or \OII emitters)
at $z=1$ that are generated with our method using TNG300-D and Large simulations.
For comparison, we also show the power spectra of matter in TNG300-D and Large simulations,
mock LRGs in Large simulation\footnote{We do not measure the power spectrum of mock LRGs in TNG300-D
because there are not enough mock LRGs in TNG300-D simulation due to the small simulation volume.} and
the result of \textit{halofit} fitting formula \citep{Smith2003} with refined parameters by \citet{Takahashi2012}.
The power spectra of mock ELGs based on our method fairly well reproduce the power spectra of the
\HA or \OII emitters identified in the hydrodynamical simulation over the overlapping range of scales.
The ratio between power spectra of matter and ELGs is almost constant at large scales
but has weak scale dependence at small scales.

\section{Results}
\label{sec:results}
Here, we present the properties of mock ELGs populated by the new scheme in more detail.
We also discuss how well the mock ELGs reproduce the results of the hydrodynamical simulation.
In this section, all of results are based on the snapshot at $z = 1$.

\subsection{Radial profile in halos}
\label{sec:radial_profile}
\begin{figure}
\includegraphics[width=\columnwidth]{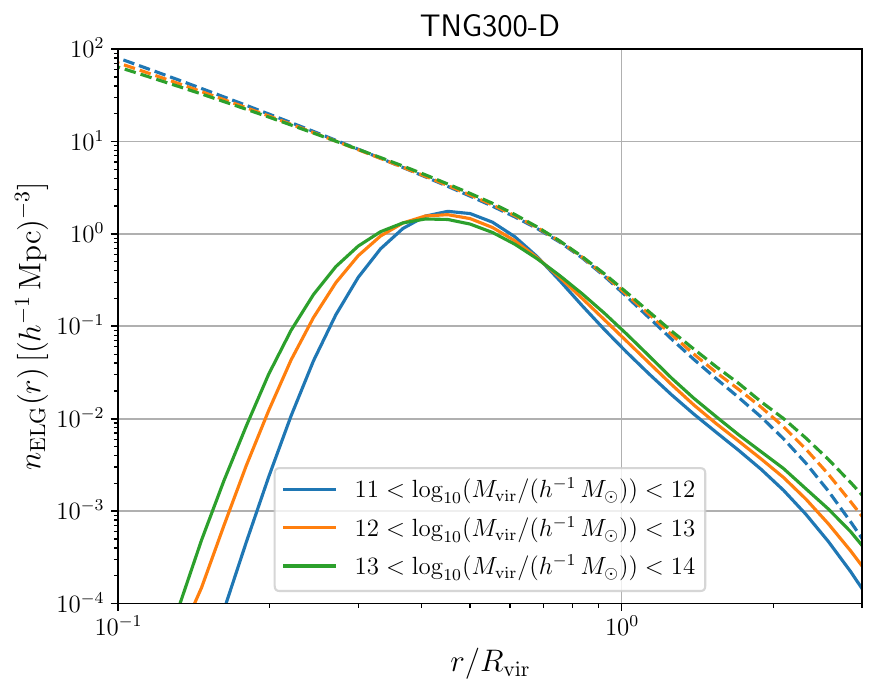}
\includegraphics[width=\columnwidth]{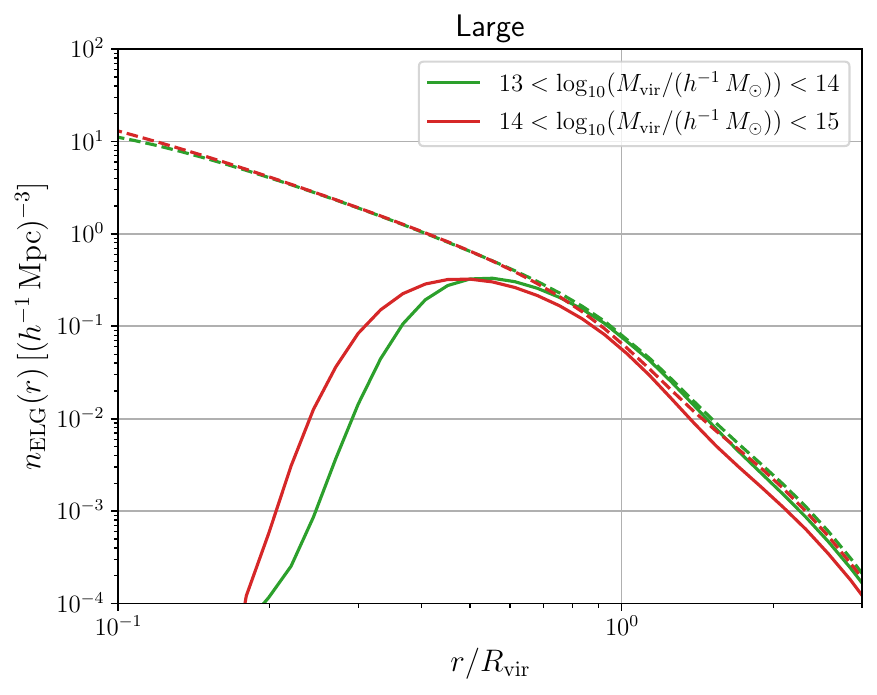}
\caption{The radial profiles of ELGs at redshift $z=1$ for mass bins:
$M_\mathrm{vir}/(h^{-1} \, \Msun) \in [10^{11}, 10^{12}], [10^{12}, 10^{13}], [10^{13}, 10^{14}]$ for TNG300-D (upper panel)
and $M_\mathrm{vir} / (h^{-1} \, \Msun) \in [10^{13}, 10^{14}], [10^{14}, 10^{15}]$ for Large (lower panel).
The profiles are normalised to match
the number density $\bar{n}_\mathrm{ELG} = 10^{-3} \, (\hMpc)^{-3}$ by introducing the constant weight (Eq.~\ref{eq:weight}).
The radius is normalised by the virial radius $R_\mathrm{vir}$.
The dashed lines show profiles for all particles with the same weight, which are proportional to
matter density profile.}
\label{fig:profile}
\end{figure}

First, we measure the average radial number density profile of ELGs
in host halos, subdivided into different virial mass \citep{Bryan1998} bins:
$M_\mathrm{vir} / (h^{-1} \, \Msun) \in [10^{11}, 10^{12}],
[10^{12}, 10^{13}], [10^{13}, 10^{14}]$ for TNG300-D and
$M_\mathrm{vir} / (h^{-1} \, \Msun) \in [10^{13}, 10^{14}], [10^{14}, 10^{15}]$ for Large.
Figure~\ref{fig:profile} shows the profiles of mock ELGs measured from the simulations.
For comparison, we also show the scaled profile of all $N$-body particles,
which is equivalent to the \textit{scaled} mass density profile.
The figure shows a clear depletion of ELGs in the central region of each host halo,
as our method, more exactly the upper limit of the local density, prevents ELGs from residing
in the highest mass density region such as the central region of host halos.
On the other hand, the lower limit of density threshold has only minor impact on
the number density at the outskirts because the density is still higher than the introduced lower limit.
Some studies based on the SAM \citep{Orsi2018}, hydrodynamical simulations \citep{2021MNRAS.502.3599H},
and observations \citep{Alpaslan2016,Favole2017,Khostovan2018,Guo2019} imply
that some ELGs are found even near the centre of halos,
but in contrast, our results indicate a sharp cutoff of number density profile near the centre.
Though this feature can be considered as one of limitations of our method,
most of ELGs hosted in massive halos are satellite galaxies and
for low-mass halos, the cutoff scale in physical length is quite small
and thus our mock ELGs can be populated near the centre.
Hence we believe that this cutoff feature does not have critical impacts on the clustering properties.

\subsection{Cosmic web classifications}
\label{sec:cosmic_web}
\begin{figure}
\includegraphics[width=\columnwidth]{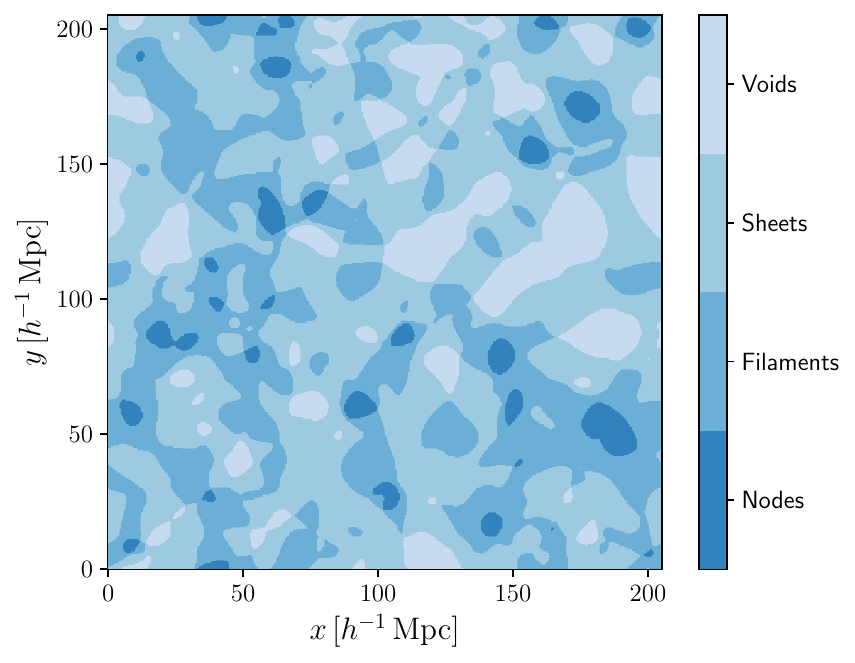}
\caption{An example of environment classification according to the tidal tensor
for the slice with thickness $0.4 \, \hMpc$ in TNG300-D at redshift $z=1$.
Based on the order of eigenvalues of the tidal tensor,
four cosmic web structures (nodes, filaments, sheets, and voids) are illustrated.}
\label{fig:web}
\end{figure}

In this section, we show the main results of this paper.
SAM predicts that ELGs are more likely to reside in filamentary structures \citep{Gonzalez-Perez2020}
because ELGs are newly formed objects and are still in the transition stage of falling towards halos at a later time.
In order to investigate the environment dependence of ELGs,
we first compute the tidal tensor of gravitational field \citep{Hahn2007a,Hahn2007b,Cautun2014,Libeskind2018}
to characterize different environments of the cosmic web in large-scale structure.
For this purpose, we use the dimensionless tidal tensor $T_{ij}$, which is defined as
\begin{equation}
  T_{ij} (\bm{x}) \equiv \frac{1}{4 \pi G \bar{\rho}_\mathrm{m}}
  \frac{\partial^2 \phi}{\partial x_i \partial x_j} ,
\end{equation}
where $G$ is the gravitational constant, $\bm{x}$ is the comoving coordinates,
and the gravitational potential $\phi$ is obtained from the Poisson equation:
\begin{equation}
  \nabla^2 \phi (\bm{x}) = 4 \pi G \bar{\rho}_\mathrm{m} \delta (\bm{x}).
\end{equation}
First, we compute the density contrast field $\delta (\bm{x})$ with cloud-in-cell (CIC) assignment
in regular grids with the number of grids on a side,
$n_\mathrm{grid} = 512 \ (1024)$,
corresponding to the grid size $L/n_\mathrm{grid} = 0.40 \ (1.95) \, \hMpc$ for TNG300-D (Large).
Next, we perform fast Fourier transform to solve the Poisson equation
and convolve the transformed field with a Gaussian filter with smoothing length $R_\mathrm{s} = 5 \, \hMpc$,
to reduce transient structures.
We also deconvolve aliasing effect due to CIC assignment \citep{Jing2005}.
Then, we compute the eigenvalues of the tidal tensor,
which are denoted as $\lambda_i \ (i = 1, 2, 3)$ in an ascending order,
and each grid position can be classified into four categories according to the following conditions:
\begin{itemize}
  \item Nodes: $\lambda_\mathrm{th} < \lambda_1 < \lambda_2 < \lambda_3$,
  \item Filaments: $\lambda_1 < \lambda_\mathrm{th} < \lambda_2 < \lambda_3$,
  \item Sheets: $\lambda_1 < \lambda_2 < \lambda_\mathrm{th} < \lambda_3$,
  \item Voids: $\lambda_1 < \lambda_2 < \lambda_3 < \lambda_\mathrm{th}$.
\end{itemize}
For the threshold value, we adopt the fiducial threshold $\lambda_\mathrm{th} = 0.01$,
which reproduces well the visual impression of cosmic web classification \citep{ForeroRomero2009,Cui2018}.
We have checked that the classification process is less subject
to choice of the smoothing length and the threshold value.
Figure~\ref{fig:web} visualizes
environmental categories in the cosmic web
for a slice of TNG300-D simulation \citep[also see][]{Cautun2014}.
It is clear that sheets and voids give a dominant volume fraction of large scale structure,
while nodes including halos have a tiny volume fraction.
However, nodes give a decent \textit{mass} fraction in the cosmic web,
while voids give the smallest mass fraction.
Since galaxies tend to live in halos, galaxies in massive halos affect clustering properties
as seen from those of luminous red galaxies.

\begin{figure}
\includegraphics[width=\columnwidth]{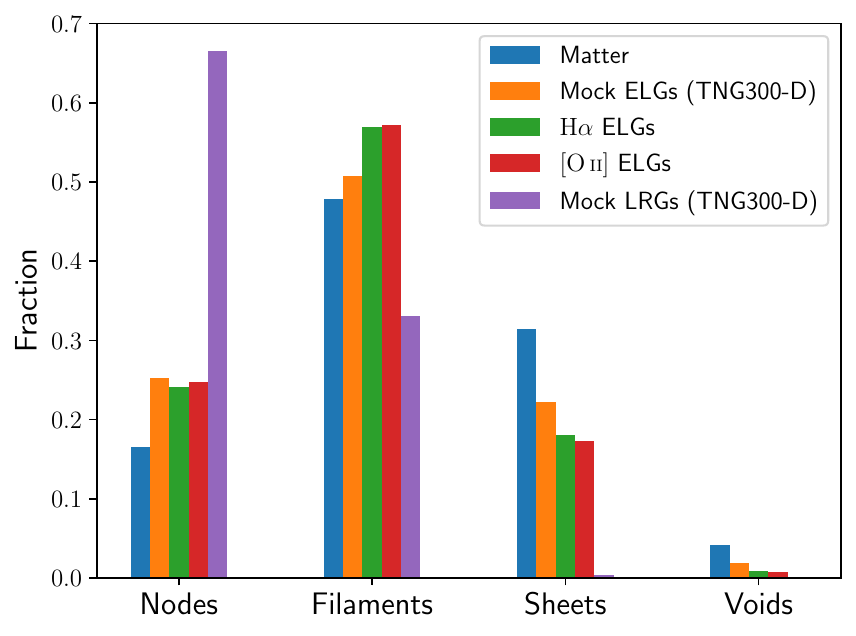}
\caption{The fraction of four classes (nodes, filaments, sheets, and voids) at redshift $z=1$
for different samples: matter, mock ELGs, mock LRGs, and \HA and \OII ELGs taken
from full physics hydrodynamical simulation (TNG300).}
\label{fig:class}
\end{figure}

Figure~\ref{fig:class} shows the most important result of this paper,
which studies in which environment of the cosmic web, i.e. nodes, filaments, sheets or voids,
mock galaxies tend to reside. Here we assign, to each galaxy or object, the environment at the nearest grid.
For comparison, we also show the results for matter and the mock LRG sample.
Encouragingly, our mock ELGs show
a similar tendency in their environments to that of
ELGs from hydrodynamical simulations;
about half of ELGs reside in filaments, the remaining quarters are located in nodes and sheets,
and the rest,
a considerably small fraction, are in voids.
On the other hand, matter is more likely to be found in sheets and
most of mock LRGs are found in nodes as expected.
Thus, our mock ELGs fairly well reproduce the environment dependence of ELGs as seen in hydrodynamical simulations.

\subsection{Dependence of density thresholds}
\label{sec:thresholds}
\begin{figure}
\includegraphics[width=\columnwidth]{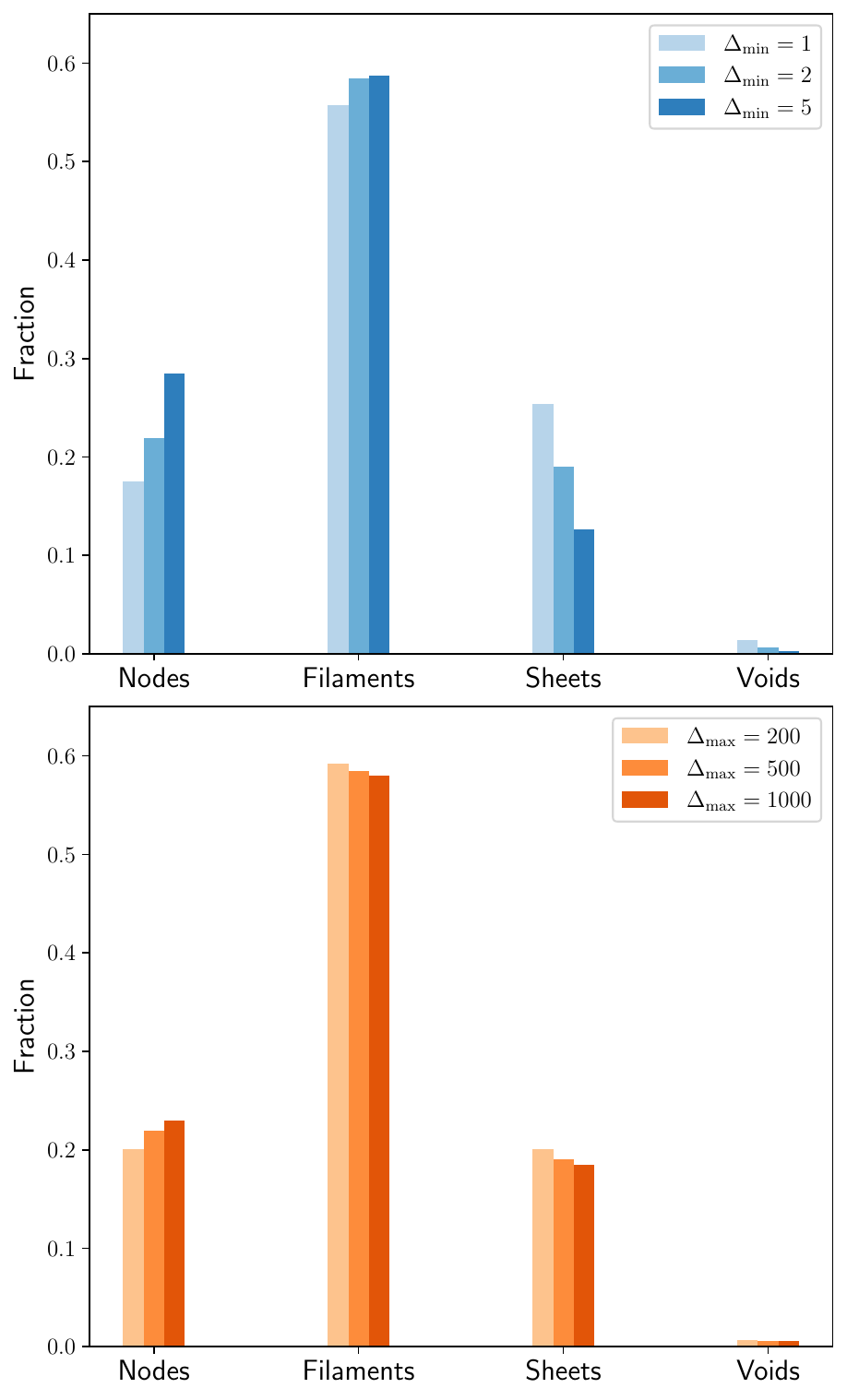}
\caption{The dependence of environment fraction on the density thresholds for Large simulation
at redshift $z=1$.
In upper (lower) panel, we vary $\Delta_\mathrm{min}$ ($\Delta_\mathrm{max}$)
and fix $\Delta_\mathrm{max} = 500$ ($\Delta_\mathrm{min} = 2$).}
\label{fig:class_limit}
\end{figure}

In principle, the density threshold is arbitrary but adjusted so that
the obtained ELGs satisfy the desired property.
In this work, we match the large-scale bias and environment fractions
by tuning two thresholds: $\Delta_\mathrm{min}$ and $\Delta_\mathrm{max}$.
Figure~\ref{fig:class_limit} shows how the environment fraction varies
with different choices of the
minimum or maximum threshold for Large simulations.
As an overall trend, the minimum threshold has more impacts on the large-scale bias \citep{Pujol2017}.
On the other hand, the maximum threshold is more sensitive
to the small-scale clustering (see the next section).

\begin{figure*}
\includegraphics[width=0.9\textwidth]{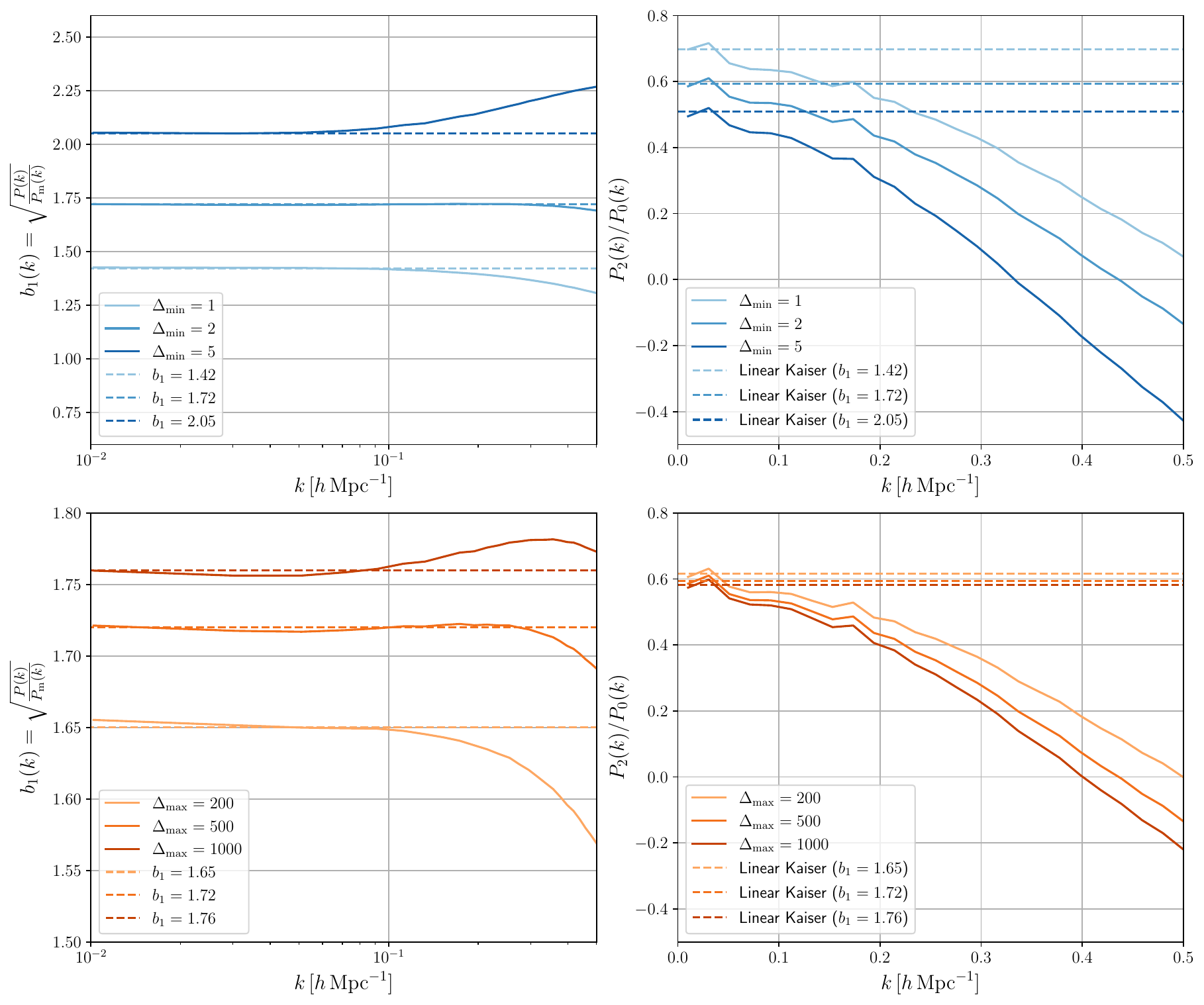}
\caption{The dependence of power spectra on the density thresholds for Large simulation
at redshift $z=1$.
In upper (lower) panels, we vary $\Delta_\mathrm{min}$ ($\Delta_\mathrm{max}$)
and fix $\Delta_\mathrm{max} = 500$ ($\Delta_\mathrm{min} = 2$).
In left and right panels, the scale-dependent bias and
the ratio between quadrupole and monopole moments are shown, respectively.
In the left panels, the dashed lines correspond to the constant bias cases,
and in the right panels, the linear Kaiser predictions (Eq.~\ref{eq:linear_Kaiser}) with the constant bias are shown.}
\label{fig:power_limit}
\end{figure*}

\subsection{Real- and redshift-space power spectra}
\label{sec:power_spectrum}

In this section, we study the real- and redshift-space power spectra of mock ELGs and focus only on the Large simulation
because the small volume ($0.009 \, (h^{-1} \, \mathrm{Gpc})^3$) covered by TNG300-D
is not comparable with the realistic galaxy clustering measurements.
The left panels of Figure~\ref{fig:power_limit} show the ratio of the real-space power spectrum of ELGs to that of matter,
which gives the effective bias function defined as
$b_1 (k) \equiv \sqrt{P(k)/P_\mathrm{m}(k)}$.
Here we use all the candidate $N$-body particles of mock ELGs satisfying
the target density range of $\Delta_\mathrm{min} < \rho_i/\bar{\rho}_\mathrm{m} < \Delta_\mathrm{max}$,
so the shot noise contamination is negligible.
The figure shows that the bias approaches to a scale-independent value,
the linear bias parameter, at the limit of $k \rightarrow 0$, and the value depends on
the choice of the density threshold.

\begin{figure}
\includegraphics[width=\columnwidth]{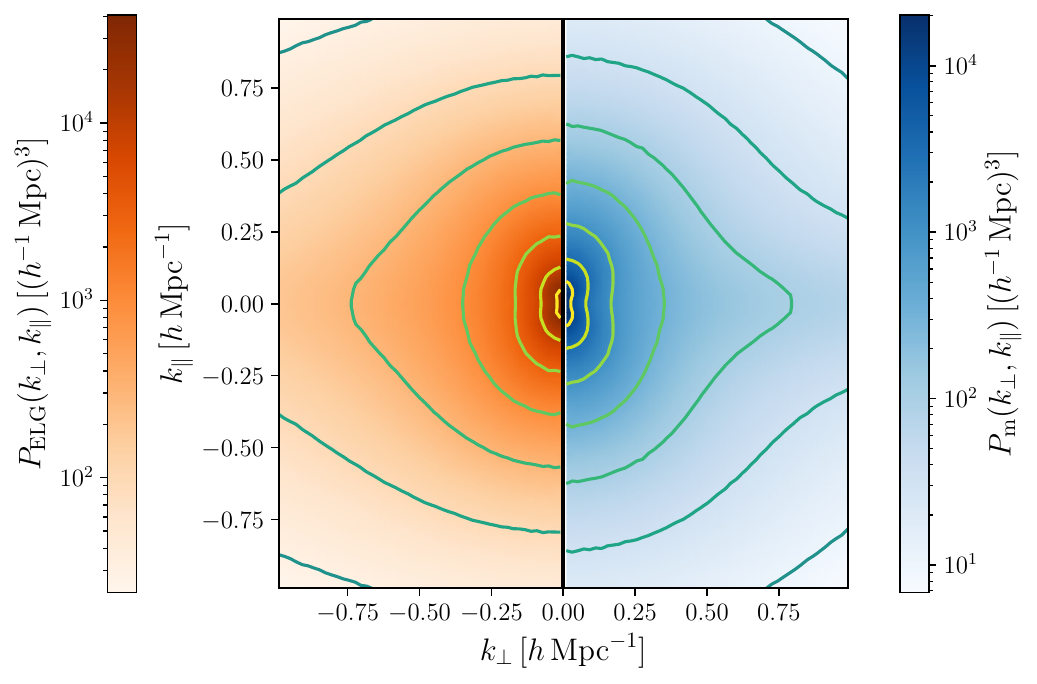}
\includegraphics[width=\columnwidth]{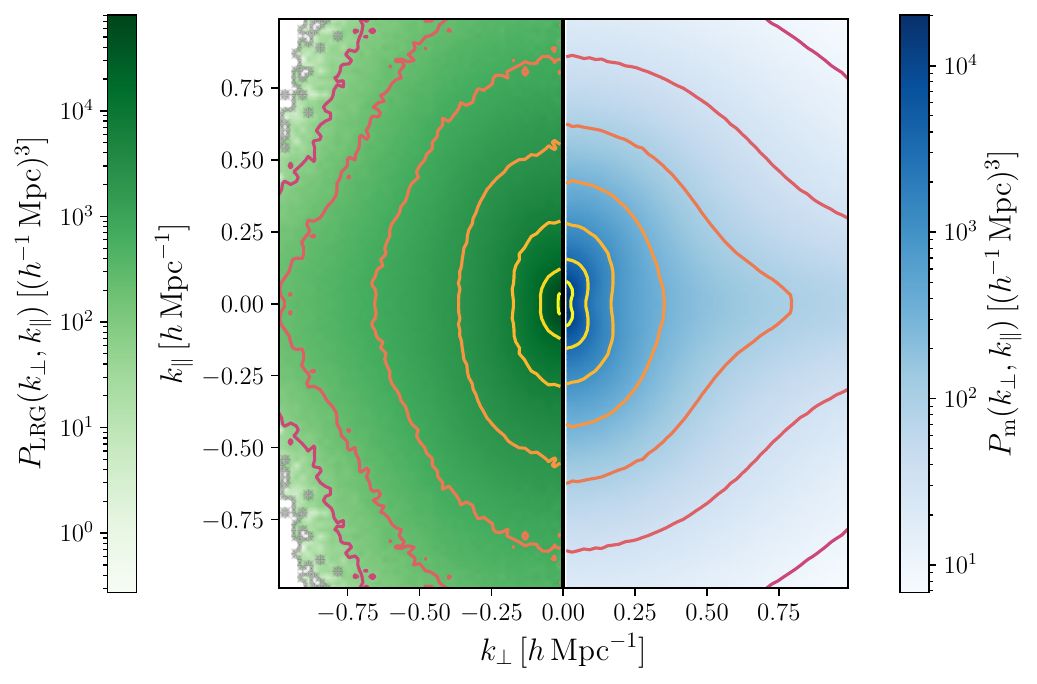}
\caption{The 2D anisotropic power spectra of matter (upper and lower right panel),
mock ELGs (uppper left panel), and mock LRGs (lower left panel)
in Large simulation at redshift $z=1$.
The contours are drawn in the range
$[10^1, 10^4] \, (\hMpc)^3$ with 7 log-equally spaced levels
for matter power spectrum. For ELG and LRG power spectra,
the contours are drawn in the range scaled with bias
$[b_1^2 10^1, b_1^2 10^4] \, (\hMpc)^3$ ($b_1 = 1.72$ for mock ELGs
and $b_1 = 2.64$ for mock LRGs)
with the same spacing.}
\label{fig:pk2D}
\end{figure}

Now we study the redshift-space power spectrum of mock ELGs,
which is one of the most important observables for
cosmology based on galaxy clustering.
The redshift-space distortion (RSD) effect due to peculiar velocities of ELGs causes an apparent anisotropic clustering
in the redshift-space galaxy distribution.
As a result, the redshift-space power spectrum is generally given as a function of two wave-numbers, $k_\parallel$ and $k_\perp$,
which are the components of wave-numbers parallel and perpendicular to the line-of-sight direction, respectively.
Since ELGs preferentially reside in filaments, the RSD effect might be different from that of massive subhalos hosting LRGs
that have been well studied in the literature.
Figure~\ref{fig:pk2D} shows the redshift-space power spectra of matter, mock ELGs, and mock LRGs.
The redshift-space power spectrum of ELGs displays a weaker stretching effect of the clustering power
along the direction parallel to the line-of-sight direction on large wavenumbers,
compared to that of the matter power spectrum. The stretching effect is caused by the random motions of ELGs, i.e. FoG effect.
This result is as expected because ELGs would tend to reside in smaller halos,
which have a smaller velocity dispersion, and in addition
avoid the central regions of massive halos, which have
a larger velocity dispersion.
In contrast, mock LRGs sample at $z=1$ show the weakest FoG feature because
most of such massive subhalos are central and thus the velocity dispersion is quite small. Note
that LRGs at the lower redshifts such as those of the SDSS survey display a significant FoG feature
because the sample contains more satellite LRGs in massive halos
\citep{2012MNRAS.426.2719R}. Clustering properties of LRGs at $z\sim 1$ have yet to be understood and
need to be carefully studied from actual data, e.g., DESI.

On the other hand, the large-scale RSD effect (i.e. the effect on small wavenumbers) arises
mainly from the coherent large-scale peculiar velocity field of ELGs \citep{Kaiser1987}.
Since ELGs preferentially reside in filaments,
the large-scale RSD effect of ELGs might be different from that of LRGs that have been well studied
in the literature such as the SDSS studies.
To quantify the large-scale RSD effect, we use the multipole moments of the redshift-space power spectrum, defined as
\begin{equation}
  P_\ell (k) = \frac{2 \ell + 1}{2} \int_{-1}^{+1} P(k, \mu) L_\ell (\mu) \mathrm{d}\mu ,
\end{equation}
where $L_\ell (x)$ is the $\ell$-th order Legendre polynomial.
The linear theory gives the Kaiser formula for the monopole and quadrupole moments \citep{Kaiser1987}:
\begin{align}
  P_0 (k) = & \left( b_1^2 + \frac{2}{3} f b_1 + \frac{1}{5} f^2 \right) P_\mathrm{lin} (k) , \\
  P_2 (k) = & \left( \frac{4}{3} f b_1 + \frac{4}{7} f^2 \right) P_\mathrm{lin} (k) ,
\end{align}
where $f \equiv \mathrm{d} \ln D_+ / \mathrm{d} \ln a$ is the linear growth rate,
$D_+$ is the linear growth factor, and $P_\mathrm{lin} (k)$ is the linear matter power spectrum.
Thus, the linear theory predicts the ratio
\begin{equation}
  \label{eq:linear_Kaiser}
  P_2(k)/P_0(k) = \left. \left( b_1^2 + \frac{2}{3} f b_1 + \frac{1}{5} f^2 \right)
  \middle/ \left( \frac{4}{3} f b_1 + \frac{4}{7} f^2 \right) \right. ,
\end{equation}
i.e. the scale-independent value that depends only on the linear bias and the growth rate.

\begin{figure}
\includegraphics[width=\columnwidth]{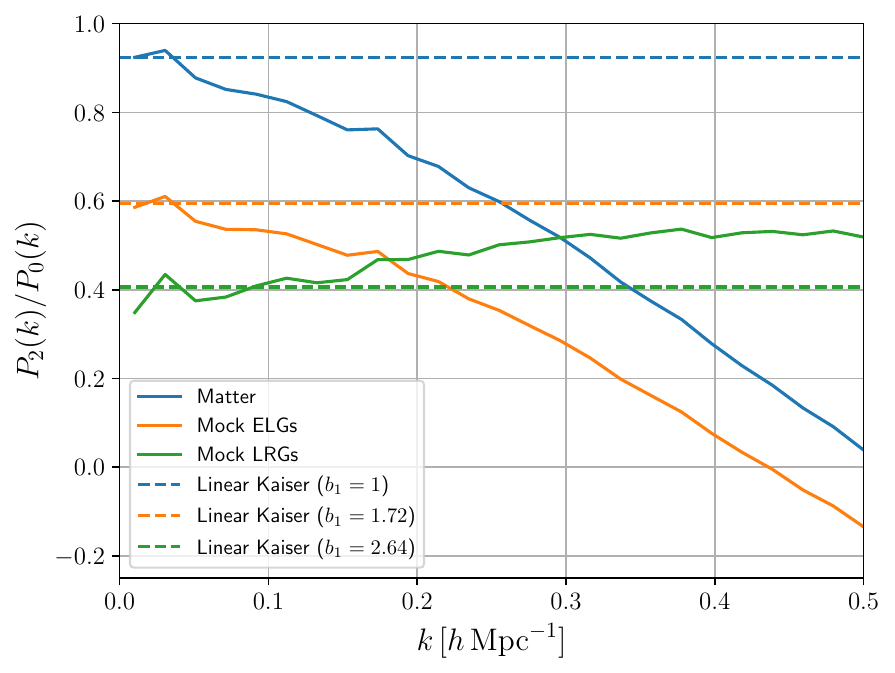}
\caption{The ratio between quadrupole and monopole moments for matter, mock ELGs,
and mock LRGs at redshift $z=1$.
The dashed line show the result of linear Kaiser formula
with linear bias $b_1 = 1$ (matter), $1.72$ (mock ELGs), and $2.64$ (mock LRGs).}
\label{fig:multi_ratio}
\end{figure}

In Figure~\ref{fig:multi_ratio}, we show the ratio of the monopole and quadrupole moments for matter, mock ELGs, and mock LRGs.
The figure shows that the ratio for ELGs has a different $k$-dependence from that of matter and mock LRGs, which is partly due to the weaker
FoG effect as discussed in Figure~\ref{fig:pk2D}.
Encouragingly the ratio at the small $k$ limit approaches the Kaiser formula prediction (dashed line),
suggesting that the large-scale ratio should be used to extract the growth rate information as done for the LRG power spectrum.
The right panels of Figure~\ref{fig:power_limit} show how the ratio varies with different choices of
the density thresholds $\Delta_\mathrm{min}$ and $\Delta_\mathrm{max}$ in our method.
For all cases, the ratio is given by the Kaiser formula using the linear bias value for each sample.

To further gain physical insight on the large-scale RSD effect of mock ELGs,
we study the cross-power spectrum of the \textit{real-space} matter field with the \textit{redshift-space} density field of ELGs
in different environments. This is not observable, but we can find some interesting features from this statistics.
Using the linear theory, we can find that the cross-power spectrum, denoted as $P^{\mathrm{m}^r \mathrm{ELG}^s}(k, \mu)$,
is given as
\begin{equation}
  P^{\mathrm{m}^r \mathrm{ELG}^s}(k, \mu) = \left( b_1 + f \mu^2 \right) P_\mathrm{lin} (k) .
\end{equation}
Then, the quadrupole moment of this cross-power spectrum is given as
\begin{equation}
  \label{eq:linear_Kaiser_cross}
  P_2^{\mathrm{m}^r \mathrm{ELG}^s} (k) = \frac{2}{3} f P_\mathrm{lin} (k) .
\end{equation}
That is, the quadrupole moment amplitude relative to the linear matter power spectrum does not depend on the linear bias,
and only depends on the growth rate.
Motivated by this prediction, we use the quadrupole moment of the cross-power spectrum measured from the mock ELG catalog,
and then study the ratio to the matter spectrum measured from the same simulation,
$P_2^{\mathrm{m}^r \mathrm{ELG}^s} (k)/P_\mathrm{m}(k)$.

\begin{figure}
\includegraphics[width=\columnwidth]{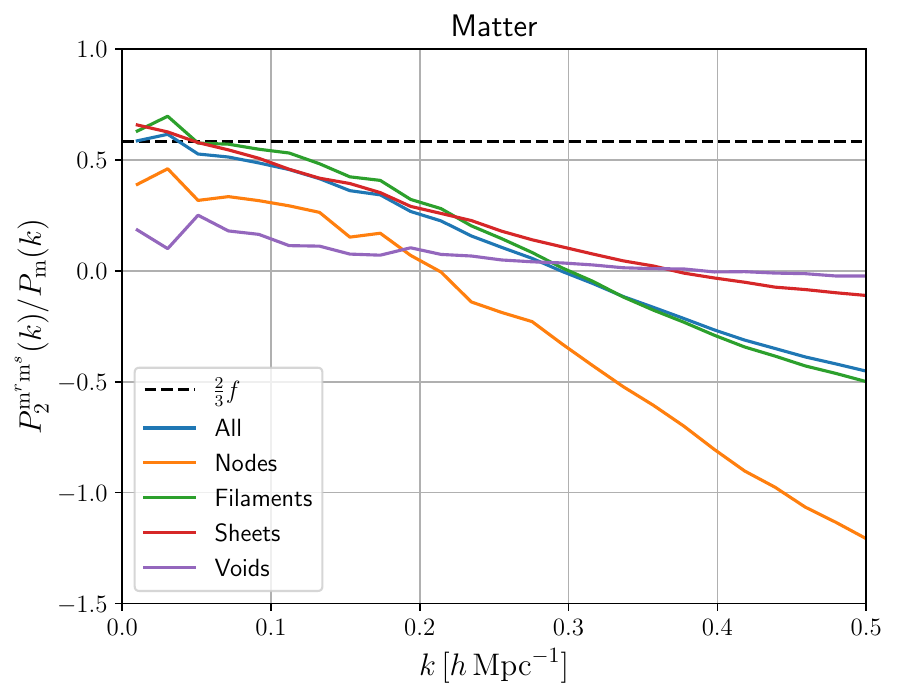}
\includegraphics[width=\columnwidth]{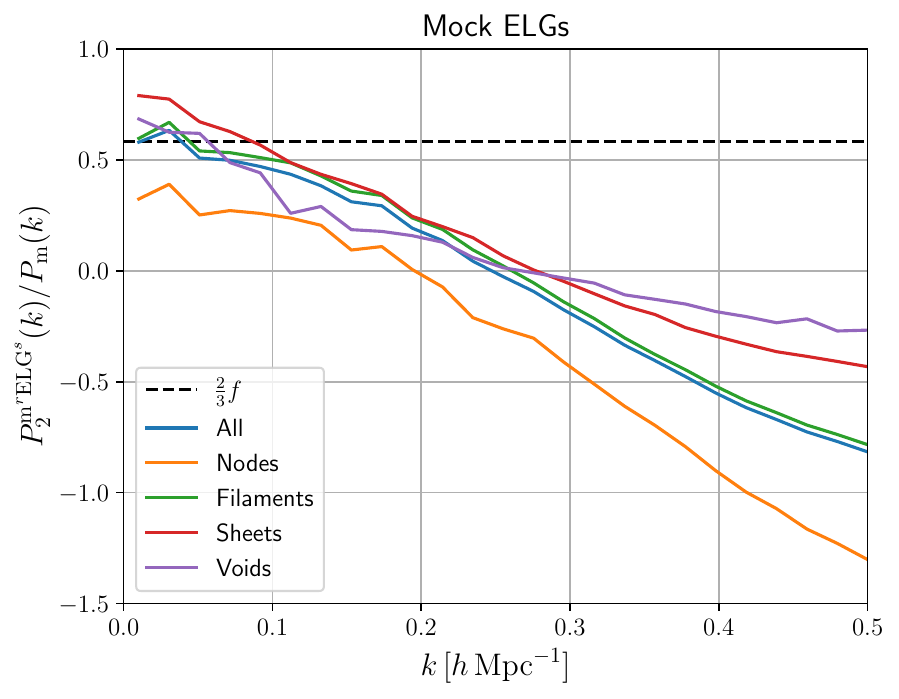}
\includegraphics[width=\columnwidth]{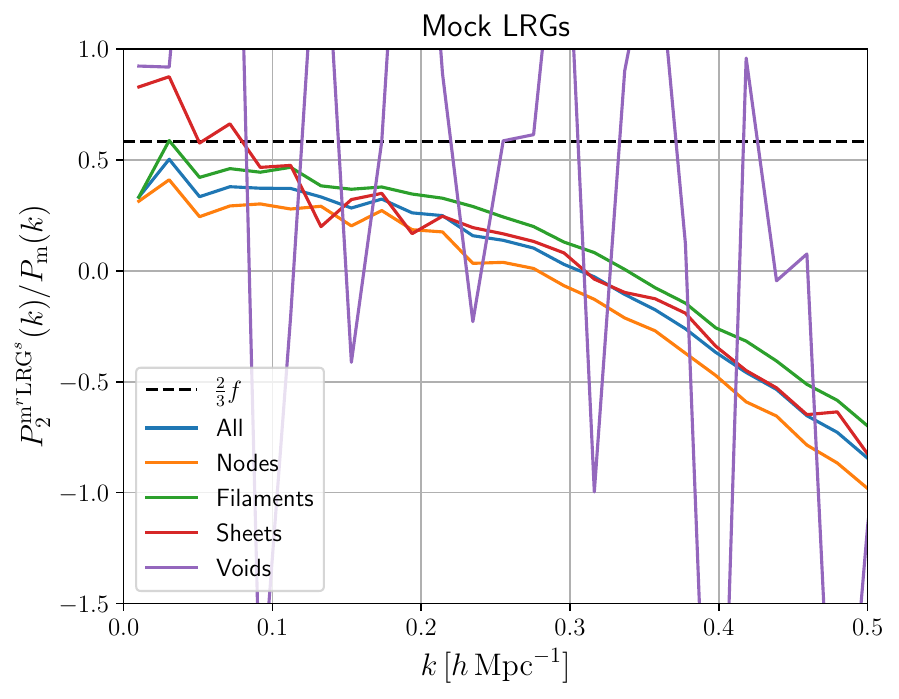}
\caption{The ratio between quadrupole moment of real-space matter density and
redshift-space number density of matter (upper panel), mock ELGs (medium panel),
mock LRGs (lower panel) and cross-power spectra
and matter power spectrum at redshift $z=1$.
These samples are divided according to its environment and different lines correspond
to the results of four different environment (nodes, filaments, sheets, and void) or all environments.
The black dashed line corresponds to the linear prediction.
The cross-power spectrum of mock LRGs at void region is noisy
because only a few LRGs are found in void region.}
\label{fig:cross}
\end{figure}

Figure~\ref{fig:cross} shows the results where we use the cross-power spectra of matter,
mock ELGs, and mock LRGs in different environments (see Figure~\ref{fig:class}).
The ratios for different ELG samples display different $k$-dependence.
Intriguingly, the ratios for ELGs in filaments and sheets show greater amplitudes than predicted by the linear theory,
while ELGs in nodes show a smaller amplitude \citep[also see][for a similar discussion]{2020PhRvD.101b3510K}.
The similar tendency can be seen for matter and mock LRGs.
Although the large-scale RSD effect for the overall ELG sample
is close to the linear theory prediction,
different samples of ELGs or inhomogeneous selection of ELGs might give a non-trivial RSD effect
depending on which ELGs in the cosmic web to use for the clustering analysis.
In this sense, we would like to stress that it is encouraging that the fraction of ELGs
in different environments can be reproduced using a simple scheme.
The nontrivial large-scale limit of the anisotropic signal from galaxies
in different environments needs to be kept in mind.

\section{Conclusions}
\label{sec:conclusions}
ELGs are the main targets of upcoming spectroscopic surveys
to measure the large-scale structures of the Universe.
However, modelling of ELG distribution must resort to computationally expensive methods:
SAMs or hydrodynamical simulations.
Since the strong emission line of ELGs is emitted by short-lived massive stars,
the hosting halo sample is not always a
mass-limited sample, unlike
LRGs.
Furthermore, the typical mass of halos hosting ELGs is less massive, e.g. $10^{12} \, h^{-1} \, \Msun$,
and thus, high mass resolution to identify such low-mass halos is an obstacle to run large simulations.
In this work, we develop
a fast scheme to populate the ELGs into dark-matter only simulations.
This method populates ELGs onto particles
whose local density lies within the preset range.
The local density is estimated with the SPH kernel and
it is readily available in $N$-body simulations.

In this method, the particles which satisfy the local density condition are regarded as ELGs.
We have introduced two density thresholds, the minimum and maximum.
The minimum threshold makes particles evade
low density regions
and the maximum threshold avoids central regions of massive halos
because in general, galaxies found near the center are already quenched.
Although our method utilises
such a simple criterion,
the resultant ELG sample exhibits environment fraction
similar to that of ELGs simulated in hydrodynamical simulations;
only $20\%$ of ELGs reside in nodes, the rest of ELGs is in filaments and sheets,
and ELGs in voids are quite rare.
The large-scale bias and the environment fraction depends on the two density thresholds,
and thus, these thresholds should be adjusted so that the obtained sample has a desired bias and
small-scale anisotropic power spectrum, i.e. FoG effect.
Once the bias is adjusted to a reasonable range expected for ELGs,
the obtained environment fraction is consistent with hydrodynamical simulation results.
The anisotropic power spectrum suggests that ELGs should be less subject to the FoG effect
because they are confined to outer regions of halos and have less virial random velocity.
This behaviour is regulated by the maximum density threshold.
Furthermore, we have measured the cross-power spectrum of real-space matter density field and redshift-space
number density fields of ELGs in each environment.
At large scales, the ratio between the quadrupole
of the cross-power spectrum and the real-space matter power spectrum is proportional to the growth rate, i.e.,
the relation between density and velocity fields, \textit{without} the dependence on the linear bias parameter, and thus,
the ratio directly corresponds to the strength of the redshift space distortion effect.
The clear difference
on the environment of ELGs can be seen,
which is contradictory to the linear theory prediction, and it suggests that the ELGs
in filaments or sheets should be in the course of infall.
As a result, redshift space distortion of ELGs is quite different from that of matter or LRGs.

As a result, our method realises fast production of many realisations of mock ELG catalogues.
This suite of mock catalogues has multiple use for spectroscopic surveys:
estimating covariance matrix of ELG power spectra,
calibrating the analysis pipeline, and quantifying various observational systematics.
Furthermore, our method directly populates ELGs on particles, not on halos.
Thus, our method is independent from halo finding algorithm
and has an advantage over approaches which relies on halos \citep[e.g.,][]{Alam2019,Avila2020}.

\section*{Acknowledgements}
KO is supported by JSPS Overseas Research Fellowships and JSPS Research Fellowships for Young Scientists.
This work was supported in part by
Grant-in-Aid for JSPS Fellows
Grant Number JP21J00011, World Premier International Research Center Initiative (WPI Initiative), MEXT, Japan,
and JSPS KAKENHI Grant Numbers JP18H04350, JP18H04358, JP19H00677, JP20J22055, JP20H05850, JP20H05855, JP20H05861 and JP21H01081;
and by the Basic Research Grant (Super AI) of Institute for AI and Beyond of the University of Tokyo.
We also acknowledge financial support from Japan Science and Technology Agency (JST) AIP Acceleration Research Grant Number JP20317829.
The numerical calculations were carried out on Yukawa-21 at Yukawa Institute for Theoretical Physics in Kyoto University.

\section*{Data Availability}
The simulation data of IllustrisTNG is available at \url{https://www.tng-project.org}.
Other simulation data is available upon request.


\bibliographystyle{mnras}
\bibliography{main}





\bsp
\label{lastpage}
\end{document}